\documentclass[pdflatex,sn-mathphys-num]{sn-jnl}

\usepackage{url}

\usepackage[utf8]{inputenc}

\usepackage{hyperref}

\usepackage{graphicx}%
\usepackage{multirow}%
\usepackage{amsmath,amssymb,amsfonts}%
\usepackage{amsthm}%
\usepackage{mathrsfs}%
\usepackage[title]{appendix}%
\usepackage{xcolor}%
\usepackage{textcomp}%
\usepackage{manyfoot}%
\usepackage{booktabs}%
\usepackage{algorithm}%
\usepackage{algorithmicx}%
\usepackage{algpseudocode}%
\usepackage{listings}%

\usepackage{tabularx}

\usepackage{amsmath}
\usepackage{amssymb}
\usepackage{float}
\usepackage{booktabs} \usepackage{tabularx} \usepackage{array}
\usepackage{algorithm}
\usepackage{pgfplots}
\pgfplotsset{compat=1.18}
\usepackage{algpseudocode}
\usepackage{longtable}
\usepackage{booktabs} 
\usepackage{tabularx}
\usepackage{pdflscape}
\usepackage{lscape}
\usepackage{amsmath}
\usepackage{amsmath}
\usepackage{tcolorbox}
\usepackage{enumitem}
\usepackage[english]{babel}
\usepackage{amsthm}
\usepackage{graphicx}   
\usepackage{adjustbox}  
\usepackage{booktabs}
\theoremstyle{definition} 

\usepackage{tabularx}

\usepackage{graphicx} 
\usepackage{float}

\usepackage{float}
\usepackage{amsmath}

\usepackage{graphicx}
\usepackage{amsmath}
\usepackage{booktabs}

\usepackage{url}
\usepackage{hyperref}

\theoremstyle{thmstyleone}%
\theoremstyle{thmstyletwo}%

\theoremstyle{thmstylethree}%

\DeclareUnicodeCharacter{0327}{\c{}}

\begin{document}

\title[Article Title]{Statistical and Optimization Models for Generating Electricity in Energy Markets}


\author*[]{\fnm{Trisha} \sur{Lawrence}}\email{trisha.lawrence@ucalgary.ca}


\abstract{ Electricity generation has become a remunerative avenue for new investments and particularly in the area of renewable energy resources for electricity generation.
This paper introduces a 
 framework for integrating statistical methods with optimization modeling to improve decision support in electricity generation. The analysis begins by evaluating the success of mini‑grid projects installed across the globe using a probit and a Bayesian hierarchical model. Using both a probit and a Bayesian hierarchical model we estimate the probability of a mini-grid project being successful. Furthermore, we  provide a multi-resource optimization framework for determining the
optimal profit in electricity generation while
reducing carbon emissions
 in electricity production facilities.
 We analyze the impact of a battery energy storage system and the bidding behavior in the Alberta electricity market by constructing a mixed‑integer self‑scheduling model that determines optimal charging and discharging strategies. Numerical experiments show how storage systems can enhance energy arbitrage profitability and influence market clearing prices. Finally, we propose a Stochastic Dynamic Programming (SDP) model to maximize profit in electricity generation while also accounting for the volatility of prices. 
Integrating statistical methods with optimization-based modeling reveals great value in combining these methods towards achieving more effective, profitable, clean energy systems.  }




\maketitle

\section{Introduction}\label{sec1}

The availability of sustainable and clean energy in the world is one of society’s most important challenges. 
According to Ember \cite{em2026}, solar
power alone met 75\% of the net increase in electricity demand in 2026 and together with
wind, these two sources met almost all (99\%) of the global demand growth. 
In Canada, the reliability and affordability of energy is needed for the development of a robust Canadian economy - for example, energy costs are the third largest cost for mining operation (see for example \cite{NIR}). To increase availability of energy and reduce costs, energy providers will need to look at alternative energy sources. Increasingly, this means producing energy from alternative sources, such as renewable energy sources, locally.

The purpose of this paper is to provide a model which will help small produces namely private or community based producers to enter the energy market by being able to predict how to produce and store energy in a way that will minimize costs and maximize profits. Large producers may also gain strategic advantage as storage systems, particularly battery storage systems (BESS) lower operating costs, reduce reliance on expensive peaking units, and improve overall system reliability, making them a powerful tool for modern energy producers navigating increasingly variable electricity markets.

Both small and large energy producers benefit from local energy generation. Local energy generation refers to the production of electrical power close to where it is consumed rather than relying on large, centralized power plants. 
 Local energy generation often relies on distributed renewable sources like solar panels and wind turbines, supported by energy storage systems and coordinated through microgrids. Microgrids operates similar to the utility grid but are
located near energy-consuming areas, thus reducing the burden on the
utility grid. Instead of connecting all the power generation sources directly to the utility grid, microgrids operate within a grid of generators and renewable energy sources.

The novelty of this research is two-fold and first entails providing a comparison of statistical and optimization methods for maximizing profit in electricity generating. Our methods utilize  fossil fuel, natural gas and renewables in generating electricity. Next we propose a  stochastic dynamic programming approach to model a battery energy storage system and maximize profit.

The statistical method implemented in this paper, begins by extending the work done by Durand and Sahinyazan \cite{DURAN} in determining factors that influence the success of  mini-grid projects installed across the globe. For their results, factors such as Ownership, particularly Community Ownership and Storage emerge as statistically significant predictors of the success of a renewable mini-grid project.
 Furthermore, we compare the results of a Probit regression model to a Bayesian hierarchical model in determining factors that influence project success. Results indicate, that the Bayesian model is suitable in capturing the underlying probability of a project being successful. 
 
Second, we explore optimization methods for maximizing profit and minimizing costs in electricity generation and propose a mixed integer programming
model towards a battery storage system. Our model decides hour-by-hour to charge/discharge in order to maximize profit, consistent with Alberta’s market participation rules. In addition to energy storage systems balancing the supply and demand of electricity, they can also reduce carbon emissions as stored renewable energy displaces fossil fuels during peak demand thus enabling reliable clean power. 
Furthermore, battery energy storage systems provide opportunities for energy arbitrage and take advantage of hourly off
and on peak electricity prices by buying and storing electricity at times when prices are low, and then selling it back to the grid when the price of electricity is greater.

 Moreover, optimization frameworks for battery energy storage systems have been proposed in \cite{SU2025, Zhang2022,Bhattachary}. 
The impact of providing energy arbitrage by energy storage
systems in various electricity markets has been shown to be both a price-maker and price-taker \cite{Walawalkar,alberta, wind, Sios} with various statistical and optimization models being proposed in \cite{Smith, Hyde, BarlowM, Ware_2019, DURAN}.

Third, we implement a dynamic programming model to maximize the profit in  electricity generation while also accounting for the volatility of prices. 
Preliminary results indicate the dynamic programming model as a promising method when compared against the
mixed-integer programming model for generating profit.

The remainder of this paper is summarized as follows. After a brief review of the related methods in Section 2, we present in Section 3, a comparison of a Probit and Bayesian mdoel in  determining factors that influence the success of a mini-grid installed across the globe. In Section 4, we discuss mathematical models for maximizing profit in electricity generation and analyze the impact of a battery energy storage problem in storing energy and minimizing costs in Section 5. In Section 6 we propose SDP as a novel approach for maximizing profit in battery energy storage systems and the potential of models for capturing price volatility. Preliminary results indicate the dynamic programming model as a promising method when compared against the mixed-integer programming model.
Finally we summarize and conclude in Sections 7 and discuss 
directions for our future work in Section 8.

\section{Related Work}\label{sec2}

Statistical and optimization models for electricity generation have been studied in various research \cite{Smith, Hyde, BarlowM, Ware_2019}. The statistical models deal in detail with explanatory variables for electricity price fluctuations whereas the finance models focus on the stochastic part of the price change. For example, in \cite{Smith} a Bayesian  regression model is employed to model the wholesale electricity market and in \cite{Hyde}, a weather model using regression analysis is presented to predict the demand for the electricity supply system. The financial models in \cite{BarlowM, Ware_2019}, 
focus on capturing the volatility of prices in the Alberta electricity market. Moreover, Barlow \cite{BarlowM} and Ware \cite{Ware_2019} propose diffusion models and Polynomial maps respectively  to model the spikes  and capture the volatility in prices. Additionally, Jarrow, Protter and Sezer \cite{Jarrow} propose a credit risk model for a class of diffusion processes where the market observes the firm’s asset value when it crosses
certain levels and  interprets it as a significant change for the firm to make a public announcement.

Optimization models for electricity generation have also been studied in \cite{HanJee, Shafieea}.
Two-stage optimization modeling frameworks are proposed in \cite{He, HarrisonJ} for generating electricity given products, such as diesel generations and renewables. In \cite{HarrisonJ}, the electricity demand is assumed to be independently and identically distributed and the optimal investment strategy is determined using a multi-dimensional news
vendor model. Moreover, an optimization framework for a price-maker energy storage system is developed in \cite{Shafieea} which involves coordinating charging and discharging of an energy storage system in
a nodal electricity market. Further optimization models for electricity markets are proposed  in the following papers \cite{Hobbs2000,Pereira2005180}. In these papers, the problem is  formulated as a bi-level optimization
problem, consisting of bidding strategy and market clearing problems in the upper and lower levels. In \cite{Hobbs2000} a single firm-model is proposed and  is formulated as a Mathematical Program
with Equilibrium Constraints (MPEC) with a parameter-dependent spatial price equilibrium problem as the inner problem. 
In \cite{Pereira2005180} a binary expansion (BE) solution to the 
strategic bidding problem under uncertainty in short-term electricity markets is presented. The BE
scheme is used to transform the products of variables in the nonlinear bidding
problem into a mixed integer linear programming formulation.

 In \cite{Bhattachary}  a multistage stochastic programming (SP) model is formulated
whose objective is to minimize the expected total energy costs
incurred within a microgrid over a finite planning horizon. In this model, the demand, renewable supply and prices are
all modeled as random variables. Furthermore, in  \cite{Zhang2022}, a   stochastic optimization configuration model for the energy storage system is constructed. The expected revenue of the optimal configuration was found to be higher than the scheme with a fixed probability distribution model. Additionally, in \cite{SU2025},  a global battery capacity optimization method for existing buildings based on the historical load is proposed for maximizing the annualized net revenue. Results show that the proposed method can effectively identify the global optimal capacity of the battery energy storage system for the buildings and caused an annual reduction in the electricity bills.

Stochastic dynamic programming have been applied to battery storage problems in various research \cite{Zhang, Gomes, Zhen, alma,Chanp}. The model in \cite{Zhang} considered the problem of operating
a battery storage unit in a home with a rooftop solar photovoltaic
(PV) system so as to minimize expected long-run electricity costs
under uncertain electricity usage, PV generation, and electricity
prices. Their model also incorporated approximate cost-to-go functions. In \cite{Gomes}  a two-stage stochastic programming
problem is formulated where  the joint
operation of wind with photovoltaic power systems has an energy storage device. The optimal bids and the energy flow in the battery are the first-stage variables and the
energy flow is the second stage variable of the problem. In \cite{Zhen}, the price arbitrage problem is formulated using
stochastic dynamic programming and real-time prices are modeled as a Markov process. Results indicate that the proposed method captures 50\% to 90\% of arbitrage profit compared to perfect price forecasts. 
Another  multi-stage stochastic programming model is proposed in \cite{alma} where the objective function comprises two terms: (1) the expected profit of the renewable energy owner and (2) the CVaR (conventional value at risk) multiplied by a weighting factor $\beta W$, which controls the degree of risk aversion of the renewable energy owner. Furthermore, in \cite{Chanp} battery storage optimization strategies are evaluated  by comparing various stochastic dual dynamic programming (SDDP)
model configurations against a linear programming (LP) benchmark mode. To the best of our knowledge, statistical and optimization methods including dynamic programming and stochastic dynamic programming have   never been applied to a battery energy storage problem as done in our case.

\section{Mini-grids in Electricity Generation}

\subsection{A Bayesian Hierarchical Model}

First we begin by extending the work done by Duand and Sahinyazan \cite{DURAN} in determining factors that influence the success of a mini-grid project. Our proposed methods includes the following Bayesian hierarchical
model with a bernoulli response.
Consider $Y_{i}$ to be the outcome 
of minigrid project $i$, and write
$Y_{i} \sim Bernouilli (p_{i})$, 
where $p_{i}$ is the probability of  
project $i$ being successful. The 
covariates are $X_{1}$ = Storage and $X_{2}$ = Ownership, in the model. The model is presented below:

\begin{equation}\label{eq:ber}
Y_i \mid p_i \sim \text{Bernoulli}(p_i), \quad i = 1, \dots, 104
\end{equation}

\[
\text{logit}(p_i) = \beta_0 + \beta_1 X_{1i} + \beta_2 X_{2i}
\]

\[
X_{1i} : \text{Ownership}, \quad X_{2i} : \text{Storage}
\]

Now $p_{i}:$
\begin{equation} \log \left(\frac{p_i}{1-p_i}\right) = \eta_i = \beta_0+\beta_1X_{1i}+\beta_2X_{2i}. \end{equation}

now the shape of the posterior is proportional to the likelihood, \begin{equation} \pi(\boldsymbol{\beta}\mid\mathbf{y}) \propto L(\boldsymbol{\beta}\mid\mathbf{y}), \end{equation} where \begin{equation} L(\boldsymbol{\beta}\mid\mathbf{y}) = \prod_{i=1}^{104} p_i^{y_i} (1-p_i)^{1-y_i}, \end{equation}
and solving for $p_{i}$ gives
\begin{equation} p_i = \frac{\exp(\eta_i)} {1+\exp(\eta_i)}. \end{equation} 

We apply a Bayesian approach to the 
model in Eq. \ref{eq:ber} and let $\beta
^T = (\beta_{0}, \beta_{1}, \beta_{2})$ be our 
parameters of interest. In specifying the conjugate prior for the Bernoulli distribution, the beta distribution, $B(\alpha, \beta)$ is utilized. If $a$ and $b$ has an equal likelihood of being observed that is, $a=1$ and $b=1$ then the beta distribution is uniform. Therefore we define the following priors for each parameter as 

\[\beta _{0}  \sim unif(0, 1) \,\,\,\,\,\,\,\,  \beta_{1} \sim unif(0, 5) ,\,\,\,\,\,\,\, \sim unif(0,5) \]

There are a number of R packages available for the posterior estimation of this Bayesian hierarchical model. In this paper we implement the RJAGS package which uses the Markov chain Monte–Carlo sampling (MCMC) algorithm \cite{Gilks}. The MCMC samples are generated by the coda function package in rjags. N.iter was set as 10,000 with 2 chains to allow for adequate convergence and mixing of the chains to provide a good approximation to the posterior.

The Metropolis Algorithm or MCMC algorithm is applied to the scenario of
flipping a coin N times and observing p heads (success or failure of a project). For a simplistic hierarchical model, the bernoulli likelihood function is applied and proceeds as follows:

\begin{enumerate}

\item Arbitrary start at an initial value of $\beta$ (in the valid range), denoted by $\beta_{current}$.

\item Generate a proposed candidate $\delta \beta \sim (\mu = 0, \sigma )$ and denote the proposed value of the parameter as $\beta_{proposed} = \beta_{current} + \delta \beta $.

\item Compute the probability of moving to the proposed value, specifically expressed as:

\item 

\begin{equation} \eta_i^{*} = \beta_0^{*} + \beta_1^{*}X_{1i} + \beta_2^{*}X_{2i}, \end{equation}

\begin{equation} \ell(\boldsymbol{\beta}^{*}) = \sum_{i=1}^{104} \left[ y_i\eta_i^{*} - \log(1+\exp(\eta_i^{*})) \right]. \end{equation}

\begin{equation} \ell(\boldsymbol{\beta}^{(t-1)}) = \sum_{i=1}^{104} \left[ y_i\eta_i^{(t-1)} - \log(1+\exp(\eta_i^{(t-1)})) \right]. \end{equation} 

\begin{equation} \alpha = \min \left\{ 1, \exp\left[ \ell(\boldsymbol{\beta}^{*}) - \ell(\boldsymbol{\beta}^{(t-1)}) \right] \right\}. \end{equation}

If the proposed value $\beta_{proposed}$, happends to land outside of the permissible bounds of $\theta$, the prior and /or likelihood is set to zero.

\item Accept the proposed parameter value if a random value is sampled from a $[0,1]$ uniform distribution  is less than the probability of moving to a new location and the proposed parameter value and tally
the current value again.

Repeat the above steps until a sufficiently representative sample has been
generated

\end{enumerate}

\subsection{Probit Regression Model }

Here, for comparison, we specify a probit model with a binary outcome $Y_i \in \{0,1\}$ as follows:

\[
p_i = \Pr(Y_i = 1 \mid X_i) = \Phi(\eta_i)
\]

where, $\Phi(\cdot)$ is the cumulative distribution function of the standard normal distribution.

\[
\Phi(\eta_i) = \int_{-\infty}^{\eta_i} \frac{1}{\sqrt{2\pi}} 
               \exp\!\left(-\frac{t^2}{2}\right) \, dt
\]

It maps any real number $\eta_i \in (-\infty, \infty)$ 
to a probability $p_i \in (0,1)$, 
as it represents the cumulative distribution function of the standard normal distribution.

\subsection{Data Analysis}

The data comprises of 104 mini-grid installations and pilot projects in rural areas across the world. The data related to each project can be found in \cite{DURAN} and is manually encoded from a vast collection of primary documentation, including scientific publications, reports and databases from government, public utilities, regulators and rural electrification agencies, grey literature such as NGO, donor and development agency report, case studies, portfolios of private developers and news articles.

\begin{table}[ht]
\centering
\caption{Description of Variables Used in the Model}
\label{tab:variables}
\begin{tabularx}{\textwidth}{lX}
\toprule
\textbf{Variable} & \textbf{Description} \\
\midrule
Project Success (Independent) & Project success is encoded as 0--1. Mini-grid infrastructure in working conditions. \\
Project Capital Cost per Capita (PCCC) (Independent) & One-time fixed costs incurred for the purchase of equipment, installation, and connection costs. \\
Ownership & Categorical variable over four mini-grid ownership models: public-owned, private-owned, public-private partnership (PPP), and community-based. \\
Technology Type & Indicator variable for each of the common mini-grid renewable technologies: solar, hydro, wind, and biomass. \\
Region & Observations from 39 countries. \\
Renewable Proportion (Renew) & Continuous variable representing the proportion of renewable sources in the total installed capacity. \\
Storage & Indicator variable signifying projects with a battery or flywheel component. \\
Demand Management System (DM) & Indicator variable included in the model. \\
System Capacity per Capita (SCPC) & Total system generation capacity installed per capita in megawatts. \\
\bottomrule
\end{tabularx}
\end{table}

\subsection{Experimental Details}

A probit regression model is fitted to the data to capture the probability of success/failure on a mini-grid project. Additionally, the probit regression model and Bayesian  models are used to predict the probit regression coefficients. A comparison of the results is provided in the Table \ref{tab:comp1} and Figure \ref{fig:comp1} below:

\begin{table}[ht]
\centering
\caption{Probit regression coefficients full Model}
\label{tab:full}
\begin{tabularx}{\textwidth}{lrrrr}
\toprule
Term & Estimate & Std. Error & z value & Pr($>|z|$) \\
\midrule
(Intercept)        & -0.3031 & 1.8127 & -0.167 & 0.8672 \\
Ownershipcommunity &  1.4984 & 0.6542 &  2.290 & 0.0220$^{*}$ \\
Ownershipppp       & -0.9197 & 0.6338 & -1.451 & 0.1468 \\
Ownershipprivate   &  0.4847 & 0.5065 &  0.957 & 0.3386 \\
Storage            &  1.0926 & 0.4509 &  2.423 & 0.0154$^{*}$ \\
DM                 & -0.4162 & 0.5421 & -0.768 & 0.4426 \\
RegionAmericas     & -0.5201 & 0.5504 & -0.945 & 0.3447 \\
RegionAsiaPacific  & -0.5758 & 0.4656 & -1.237 & 0.2162 \\
Renew              & -0.7922 & 0.7554 & -1.049 & 0.2943 \\
PCCC               &  0.1247 & 0.1129 &  1.105 & 0.2693 \\
\bottomrule
\end{tabularx}

\vspace{0.5em}
\footnotesize{Signif. codes: \quad 0 ``***'' \quad 0.001 ``**'' \quad 0.01 ``*'' \quad 0.05 ``.'' \quad 0.1 `` '' 1}
\end{table}

\begin{table}[h!]
\centering
\caption{Probit Regression Coefficients}
\begin{tabular}{lrrrr}
\hline
\textbf{Variable} & \textbf{Estimate} & \textbf{Std. Error} & \textbf{z value} & \textbf{Pr($>|z|$)} \\
\hline
Intercept          & 0.624692  & 0.332237  & 1.880  & 0.0601 \textbf{.} \\
Ownership\_community & 1.038997  & 0.544251  & 1.909  & 0.0563 \textbf{.} \\
Ownership\_ppp      & -0.705918 & 0.530820  & -1.330 & 0.1836 \\
Ownership\_private  & 0.008785  & 0.411627  & 0.021  & 0.9830 \\
Storage             &   0.787310 &  0.369150  &  2.1332  &  0.0329 \\
\hline
\end{tabular}

\vspace{0.5em}
\begin{flushleft}
\textbf{Significance codes:} \quad
\textbf{***} $p<0.001$ \quad
\textbf{**} $p<0.01$ \quad
\textbf{*} $p<0.05$ \quad
\textbf{.} $p<0.1$ \quad
\textbf{ } not significant
\end{flushleft}\label{tab:reg}
\end{table}

\begin{table}[h!]
\centering
\caption{Emperical Mean and Standard deviation for Bayesian Model} \label{tab:bay}
\begin{tabular}{lrrrr}
\hline
\textbf{Variable} & \textbf{Mean} & \textbf{SD} & \textbf{Naive SE} & \textbf{Time Series SE} \\
\hline
Intercept          & 2.1679  & 4.71838  & 0.0746041 & 0.8832710 \\
Ownership\_community & 1.3081  & 4.70988  &  0.0744697  & 0.8474668 \\
Ownership\_ppp      & -2.3204 & 4.7450 & 0.0750254 & 0.8360369 \\
Ownership\_private  &  -1.0051 &  4.73934 & 0.0749356  & 0.8750269
 \\
Storage       & 1.5752  &  0.74131   & 0.0117212  &  0.0124255 \\
\hline
\end{tabular}

\vspace{0.5em}
\begin{flushleft}
\textbf{Significance codes:} \quad
\textbf{***} $p<0.001$ \quad
\textbf{**} $p<0.01$ \quad
\textbf{*} $p<0.05$ \quad
\textbf{.} $p<0.1$ \quad
\textbf{ } not significant
\end{flushleft}\label{tab:bay}
\end{table}


\begin{longtable}{c c c}
\caption{Probit and Bayesian Predicted Probabilities}
\label{tab:comp1}\\
\hline
\textbf{Obs} & \textbf{Probit (A)} & \textbf{Bayesian (B)} \\
\hline
\endfirsthead

\hline
\textbf{Obs} & \textbf{Probit (A)} & \textbf{Bayesian (B)} \\
\hline
\endhead

\hline
\endfoot

1   & 0.7599321 & 0.7751 \\
2   & 0.9210254 & 0.9158 \\
3   & 0.9519128 & 0.9512 \\
4   & 0.7339135 & 0.7183 \\
5   & 0.7367891 & 0.7491 \\
6   & 0.9223107 & 0.9235 \\
7   & 0.9210254 & 0.9158 \\
8   & 0.9223107 & 0.9235 \\
9   & 0.9223107 & 0.9235 \\
10  & 0.9223107 & 0.9235 \\
11  & 0.9210254 & 0.9158 \\
12  & 0.4676312 & 0.4677 \\
13  & 0.7339135 & 0.7183 \\
14  & 0.7367891 & 0.7491 \\
15  & 0.9223107 & 0.9235 \\
16  & 0.9210254 & 0.9158 \\
17  & 0.9210254 & 0.9158 \\
18  & 0.9210254 & 0.9158 \\
19  & 0.9210254 & 0.9158 \\
20  & 0.9223107 & 0.9235 \\
21  & 0.7367891 & 0.7491 \\
22  & 0.9928770 & 0.9875 \\
23  & 0.7339135 & 0.7183 \\
24  & 0.9928770 & 0.9875 \\
25  & 0.9928770 & 0.9875 \\
26  & 0.9519128 & 0.9512 \\
27  & 0.9210254 & 0.9158 \\
28  & 0.9210254 & 0.9158 \\
29  & 0.9928770 & 0.9875 \\
30  & 0.7599321 & 0.7751 \\
31  & 0.7599321 & 0.7751 \\
32  & 0.9210254 & 0.9158 \\
33  & 0.9210254 & 0.9158 \\
34  & 0.9210254 & 0.9158 \\
35  & 0.9210254 & 0.9158 \\
36  & 0.9210254 & 0.9158 \\
37  & 0.9928770 & 0.9875 \\
38  & 0.9519128 & 0.9512 \\
39  & 0.9519128 & 0.9512 \\
40  & 0.9210254 & 0.9158 \\
41  & 0.9223107 & 0.9235 \\
42  & 0.9928770 & 0.9875 \\
43  & 0.7599321 & 0.7751 \\
44  & 0.9519128 & 0.9512 \\
45  & 0.9928770 & 0.9875 \\
46  & 0.9223107 & 0.9235 \\
47  & 0.9519128 & 0.9512 \\
48  & 0.9210254 & 0.9158 \\
49  & 0.4676312 & 0.4677 \\
50  & 0.9519128 & 0.9512 \\
51  & 0.7367891 & 0.7491 \\
52  & 0.9519128 & 0.9512 \\
53  & 0.9210254 & 0.9158 \\
54  & 0.9519128 & 0.9512 \\
55  & 0.7599321 & 0.7751 \\
56  & 0.7339135 & 0.7183 \\
57  & 0.7339135 & 0.7183 \\
58  & 0.7339135 & 0.7183 \\
59  & 0.7339135 & 0.7183 \\
60  & 0.7339135 & 0.7183 \\
61  & 0.7367891 & 0.7491 \\
62  & 0.7367891 & 0.7491 \\
63  & 0.9519128 & 0.9512 \\
64  & 0.7367891 & 0.7491 \\
65  & 0.7367891 & 0.7491 \\
66  & 0.7339135 & 0.7183 \\
67  & 0.7339135 & 0.7183 \\
68  & 0.9210254 & 0.9158 \\
69  & 0.9223107 & 0.9235 \\
70  & 0.7367891 & 0.7491 \\
71  & 0.7367891 & 0.7491 \\
72  & 0.7367891 & 0.7491 \\
73  & 0.7367891 & 0.7491 \\
74  & 0.7367891 & 0.7491 \\
75  & 0.7339135 & 0.7183 \\
76  & 0.7339135 & 0.7183 \\
77  & 0.7599321 & 0.7751 \\
78  & 0.9928770 & 0.9875 \\
79  & 0.9928770 & 0.9875 \\
80  & 0.9928770 & 0.9875 \\
81  & 0.4676312 & 0.4677 \\
82  & 0.9223107 & 0.9235 \\
83  & 0.7367891 & 0.7491 \\
84  & 0.9519128 & 0.9512 \\
85  & 0.9210254 & 0.9158 \\
86  & 0.9928770 & 0.9875 \\
87  & 0.9928770 & 0.9875 \\
88  & 0.9928770 & 0.9875 \\
89  & 0.9928770 & 0.9875 \\
90  & 0.9928770 & 0.9875 \\
91  & 0.9928770 & 0.9875 \\
92  & 0.9928770 & 0.9875 \\
93  & 0.9928770 & 0.9875 \\
94  & 0.9928770 & 0.9875 \\
95  & 0.9928770 & 0.9875 \\
96  & 0.9519128 & 0.9512 \\
97  & 0.7367891 & 0.7491 \\
98  & 0.9519128 & 0.9512 \\
99  & 0.9519128 & 0.9512 \\
100 & 0.9519128 & 0.9512 \\
101 & 0.7367891 & 0.7491 \\
102 & 0.7367891 & 0.7491 \\
103 & 0.9519128 & 0.9512 \\
104 & 0.9519128 & 0.9512 \\
\end{longtable}

\begin{figure}[H]
\centering
\includegraphics[width=7cm, height=6cm]{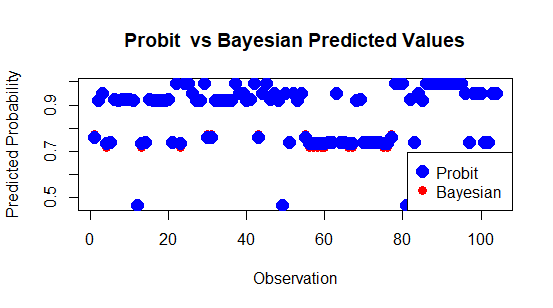} 
\caption{Plot of predictive probabilities between Bayesian and probit model }
\label{fig:comp1}
\end{figure}

\begin{figure}[H]
\centering
\includegraphics[width= 10cm]{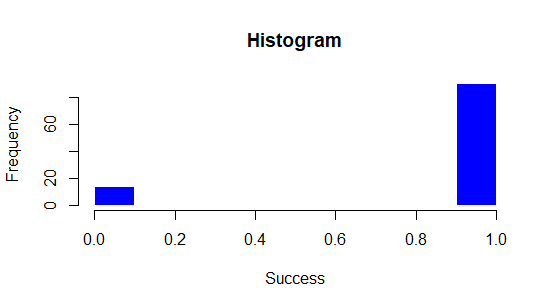}
  \caption{Histogram showing the number of successful projects}
  \label{fig:boat1}
\end{figure}

\begin{figure}[H]
\centering
\includegraphics[width= 10cm]{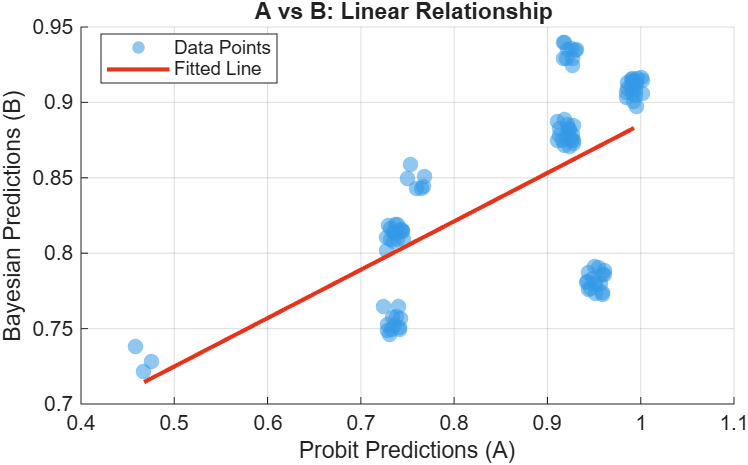}
  \caption{Scatter plot showing how closely the two models agree for each observation}
  \label{fig:scat}
\end{figure}






\subsection{Results and Discussion}

Figure \ref{fig:scat} demonstrates a scatter plot of the predicted values of the probit model and Bayesian model. The scatter points indicate how closely both models match for each observation. The fitted regression line shows the overall trend and positive linear relationship.

Table \ref{tab:full} illustrates the regression results for the probit model from the variables listed in Table \ref{tab:variables}. The model is given below as:

\[
P(\text{Success}_i = 1 \mid X_i) = \Phi(\eta_i)
\]

\[
\eta_i = \beta_0
+ \beta_1 \,\text{Private}_i
+ \beta_2 \,\text{PPP}_i
+ \beta_3 \,\text{Community}_i
+ \beta_4 \,\text{Storage}_i
+ \beta_5 \,\text{DM}_i
\]

\[
\qquad\quad
+ \beta_6 \,\text{Americas}_i
+ \beta_7 \,\text{AsiaPacific}_i
+ \beta_8 \,\text{Renew}_i
+ \beta_9 \,\text{PCCC}_i .
\]

As demonstrated in Table \ref{tab:full}, Community Ownership and Storage emerge as statistically significant predictors of project success.

Table (\ref{tab:reg}), Table (\ref{tab:bay}) and Figure (\ref{fig:comp1}) display the results of the Probit and Bayesian model considering Community Ownership and Storage. If we look into the mechanism, we will observed that the probit model adequately 
estimates the model parameters. Based on
these results, we build the following  reduced probit
regression model based on project success (Y=1).

\[
\begin{split}
\Pr(Y_i = 1 \mid X_i)
= \Phi\!\Big(
\beta_0 + \beta_1\,\text{Ownership}_{\text{community},i}
+ \beta_2\,\text{Ownership}_{\text{ppp},i}
\\
\qquad\qquad
+ \beta_3\,\text{Ownership}_{\text{private},i}
+ \beta_4\,\text{Storage}_i
\Big).
\end{split}
\]

From the probit model results we see that the probability of success increases with both Community Ownership and
Storage. Furthermore, from their  p values  0.0220 and 0.0154 Community Ownership and Storage emerge as statistically significant predictors of project success.
Thus the probit regression model is recommended as a method to determine the probability of success of a project given various factors and covariates.
\newline\indent
For the bayesian analysis, we 
ran the MCMC algorithm for 
10,0000 iterations with thin=100 (or every hundredth draw).
Convergence was assessed from 
traceplots ensuring that the 
chains mixed well. Table 
(\ref{tab:bay}) shows the posterior mean for all the 
parameters with a 95\% Bayesian credible intervals. 
The main goal is to estimate the coefficients of the model via 
simulating posterior samples. The mean posterior 
densities closely match the estimates 
determined by the probit regression model. To evaluate the predictions, we make use of the SEs as shown in Table (\ref{tab:bay}). According to Table (\ref{tab:bay}), the SE are extremely small, reflecting high precision under the bayesian framework.
\newline\indent
Table (\ref{tab:comp1}) and 
Figure (\ref{fig:comp1}) show a
comparison between the probit and Bayesian 
model. Fitting the probit 
regression model to the dataset 
using the function glm in R is a 
frequentist way of analyzing this 
dataset, while the Bayesian hierarchical linear model is 
the bayesian approach. We compare the point estimates of both 
methods via SE's and 
predictions. The naive SE is the standard error of the mean captures simulation error of the mean rather than the posterior uncertainty. While the Time Series Standard Error (TSSE) (or Monte Carlo SE) accounts for autocorrelation by using the effective sample size. The Naive SE underestimates uncertainty due to autocorrelation, whereas TSSE provides a better measure of MCMC estimation error

 Also based on the model
predictions seen in Figure (\ref{fig:comp1}), both the probit and the Bayesian models are 
suited to handling probabilities. In conclusion, the probit  
model captures the significant
relationship between the project success and its covariates, Community Ownership and Storage and its predictive 
performance is closely related to the 
Bayesian model. Bayesian methods are useful when the data is scarce and particularly for handling latent variables, in our case the probability of a mini-project being successful.

\section{Mathematical Optimization Methods in Electricity generation}\label{sec3} In this section, we provide a multi-resource optimization method for determining the optimal profit and minimizing the cost of electricity generation and transmission while reducing carbon emissions.
Mathematically, investment models, in electricity production facilities are naturally formulated as optimization models constrained by other optimization models. 
A general instance of such a model is given by the following:

\begin{align}
\max_{x,\, y^{(1)},\ldots,y^{(n)},\, \lambda^{(1)},\ldots,\lambda^{(n)}} 
\; f\!\left(x, y^{(1)}, \ldots, y^{(n)}, \lambda^{(1)}, \ldots, \lambda^{(n)}\right)
\tag{1.1a}
\end{align}

\begin{align}
\text{s.t.} \qquad 
h\!\left(x, y^{(1)}, \ldots, y^{(n)}\right) = 0
\tag{1.1b}
\end{align}

\begin{align}
g\!\left(x, y^{(1)}, \ldots, y^{(n)}\right) \le 0
\tag{1.1c}
\end{align}

\[
\left\{
\begin{aligned}
\max_{y^{(1)}}\;& f^{(1)}\!\left(x, y^{(1)}, \ldots, y^{(n)}\right) \\
\text{s.t.}\;& h^{(1)}\!\left(x, y^{(1)}, \ldots, y^{(n)}\right) = 0 \; : \; \lambda^{(1)} \\
& g^{(1)}\!\left(x, y^{(1)}, \ldots, y^{(n)}\right) \le 0
\end{aligned}
\right.
\tag{1.1d}
\]

\[
\vdots
\]

\[
\left\{
\begin{aligned}
\max_{y^{(n)}}\;& f^{(n)}\!\left(x, y^{(1)}, \ldots, y^{(n)}\right) \\
\text{s.t.}\;& h^{(n)}\!\left(x, y^{(1)}, \ldots, y^{(n)}\right) = 0 \; : \; \lambda^{(n)} \\
& g^{(n)}\!\left(x, y^{(1)}, \ldots, y^{(n)}\right) \le 0
\end{aligned}
\right.
\tag{1.1e}
\]

Problem (1.1a-1.1c) is the upper level problem while problems (1.1d) - (1.1e) are the lower level problems. The upper-level problem is the investment problem, which seeks to identify the
investment decisions that maximize profit, while the lower-level problems represent
market-clearing conditions.

\subsection{Multi-Resource Investment Problems }

Minimizing electricity generation costs while reducing carbon emissions can be formulated as a multi‑resource investment problem, where a firm must allocate 
$m$ resources to produce $n$
 products across $t$ time periods.
In this setting, the single product is electricity, which can be generated using two resources: diesel purchases and a renewable capacity generator. A two‑stage optimization model is applied where in the first stage, a nonnegative vector of capacity levels $k = (K_{A}, K_{B})$ $\in  R^{2}
+$, can be chosen representing
the diesel purchase and renewable generator capacity respectively, before observing the actual energy demand, $D \in \mathcal{R}+$,
and before the renewable energy supply is known. In the second stage, a vector of dispatch quantities $x= (x_{a}, x_{b})$  can be determined. The second stage problem with dual variables can be written as follows:

\[
C_2(K_A, K_B, \delta, \gamma) =
\min_{(x_a, x_b) \in \mathbb{R}^2_{+}, \; p \geq 0}
\; e_a x_a + c_p p
\]

subject to

\[
- x_b \;\geq\; -\gamma K_B \quad \leftarrow \lambda_b 
\]

\[
- x_a \;\geq\; -K_A \quad \leftarrow \lambda_a 
\]

\[
p + x_a + x_b \;\geq\; \delta. 
\]

The first-stage problem is defined as:

\[
\min_{(K_A, K_B) \in \mathbb{R}^2_{+}} \;
C_1(K_A, K_B)
= c^k_A K_A + c^k_B K_B
+ \mathbb{E}_{D,\Gamma}\!\left[ C_2(K_A, K_B, D, \Gamma) \right].
\]
Solving this problem involves partitioning the uncertainty space into six regions in each of which the optimal second-stage
dispatch decisions and the dual variables have the same structure.

First,  we choose a non-negative vector of capacity levels

\[
\mathbf{k} = (K_A, K_B) \in \mathbb{R}^2_+
\]

representing diesel purchase and renewable generator capacity.

The demand is a continuous nonnegative random variable written as

\[
D \in \mathbb{R}_+,
\]

and the renewable supply is realized as

\[
k_b = \Gamma K_B, \quad \Gamma \sim U[0,1].
\]

For the second stage problem,

\[
C_2(K_A, K_B, \delta, \gamma) = 
\min_{(x_a,x_b)\in \mathbb{R}^2_+, \; p \geq 0} \; e_a x_a + c_p p
\]

subject to
\begin{align}
- x_b &\geq -\gamma K_B \quad \leftarrow \lambda_b \tag{2} \\
- x_a &\geq -K_A \quad \leftarrow \lambda_a \tag{3} \\
p + x_a + x_b &\geq \delta. \tag{4}
\end{align}

and the dispatch-vector is given by:

\[
\mathbf{x} = (x_a, x_b) \in [0,K_A] \times [0,k_b],
\]

The first-stage problem is given by

\[
C_1(K_A, K_B) = c^k_A K_A + c^k_B K_B + \mathbb{E}_{D,\Gamma}\big[ C_2(K_A, K_B, D, \Gamma) \big].
\]

\vspace{25mm}

\textbf{The Demand Space is partitioned as follows:}

For \((\delta,\gamma) \in \mathbb{R}_+ \times [0,1]\), the regions are given by:

\[
\Omega_1(\mathbf{k}^*) = \{ (\delta,\gamma) \mid \delta \leq \gamma K_B \}, 
\quad \lambda_b^* = 0, \; \lambda_a^* = 0
\]

\[
\Omega_2(\mathbf{k}^*) = \{ (\delta,\gamma) \mid \gamma K_B < \delta \leq \gamma K_B + K_A \}, 
\quad \lambda_b^* = \gamma e_a, \; \lambda_a^* = 0
\]

\[
\Omega_3(\mathbf{k}^*) = \{ (\delta,\gamma) \mid \gamma K_B + K_A < \delta \}, 
\quad \lambda_b^* = \gamma c_p, \; \lambda_a^* = c_p - e_a
\]

\begin{table}[htbp]
    \centering
  \caption{Partitioning of the demand space}
    \label{tab:partitioning}
    \begin{tabular}{|l|c|r|} 
        \hline 
       Partition for $(\gamma,  \delta) \in \mathbb{R} \times [0,1]$ & $\lambda^{*}_{b} (K_{A}, K_{B}, \delta, \gamma) $&  $\lambda ^{*}_{a} (K_{A}, K_{B}, \delta, \gamma)$ \\ 
        \hline 
 $   \Omega_{1}(K^*) = {(\delta, \gamma) | \delta < \gamma K_{B}}$ & 0 &  0 \\
      $ \Omega_{2} (K^*)= { ( \delta, \gamma)}$ | $\gamma K_{B} \leq \delta \leq \gamma K_{B} + K_{A} $&$ \gamma e_{a}$ & 0 \\
        \hline 
   $ \Omega_{3}(K^*)=  ( \delta, \gamma) | \gamma K_{B} + K_{A} < \delta $&$   \gamma c_{p} $& $(c_{p} - e_{a}) $
   \\\hline  \end{tabular}
\end{table}

We consider the dual to the second stage optimization problem

\begin{equation}
\begin{aligned}
&\max -\gamma K_{\textbf{}} \lambda_{b} - K_{A} \lambda _{a}&&\\
&\text{subject to:}
-\lambda _{b} \leq e_{a}&&\\
-\lambda _{a} = 0
\end{aligned}
\end{equation}

In calculating the expectation by conditioning on the demand, of renewable supply $K_{b}(K_{B}) = \Gamma (K_{B})$, we get,

$E[\Gamma]= E[E[\Gamma|D]]$

$= \int_{-\infty}^{\infty} E[{\Gamma | D=d]P(D=d)}$

$=  \int_{-\infty}^{\infty} E[{\Gamma | D=d]f_{D}(d)dd}$

\vspace{3mm}

Using this result, Table{\ref{tab:partitioning}}  and the steps in the Harrison and Van Mieghem \cite{HarrisonJ} journal, the equation below is obtained by taking the partial derivative of the Lagrangian function w.r.t to the two capacity decision variables $K_{A}$ and $K_{B}$ with vector $K_{i}\Gamma = 0 $ for $ i \in {A, B}$ (also complementary slackness condition) of the first stage problem.

\vspace{3mm}
The KKT conditions also ensure  an optimal solution $k^{*} = (K_{A}, K_{B})$. Thus we have,


 \begin{equation} \label{mathe}
\begin{pmatrix}
0 \\ 
0
\end{pmatrix} P(\Omega_{1}(K^*)) 
+ E_{\Gamma}\!\left[
\begin{pmatrix}
\gamma e a \\ 
0
\end{pmatrix}
\Big| \Omega_{2}(K^*)
\right] P(\Omega_{2}(k)) 
+ E_{\Gamma}\!\left[
\begin{pmatrix}
\gamma c_{p} \\ 
c_{p} - e_a
\end{pmatrix}
\Big| \Omega_{3}(K^*)
\right] P(\Omega_{3}(K^*)) 
= 
\begin{pmatrix}
c^{K}_{B} - \Lambda_{B} \\ 
c_{A}^k - \Lambda_{A}
\end{pmatrix}
\end{equation}

The demand space is partitioned into three regions as shown in Table  \ref{tab:partitioning}.  For Region 1, $\Omega_1: \delta < \gamma K_{B}$, the renewables alone meet the demand, for Region 2, $\Omega_2:  \gamma K_{B} \leq \delta \leq \gamma K_{B} + K_{A}$, implies that the renewables are insufficient and diesel is also used to meet the demand and Region 3 $\gamma K_{B} + K_{A} < \delta $ indicates the share where  even 
diesel
 plus 
renewables
 together are not enough, to meet the demand. The share or probability of each region is determined by integration. For example, the probability of $\Omega_1$ can be found by considering the following:

\[
\Omega_1 = \{ (\delta, \gamma) \mid \delta \leq \gamma K_B \}, 
\quad D \sim U[0,\Delta], \quad \Gamma \sim U[0,1], \quad \text{independent}.
\]

Thus,

\[
P(\Omega_1) = P(D \leq \Gamma K_B)
= \int_0^1 P(D \leq \gamma K_B)\, d\gamma
= \int_0^1 \min\!\left(1, \frac{\gamma K_B}{\Delta}\right)\, d\gamma.
\]

\textbf{Case $K_B \leq \Delta$:}

\[
P(\Omega_1) = \int_0^1 \frac{\gamma K_B}{\Delta}\, d\gamma
= \frac{K_B}{\Delta} \cdot \frac{1}{2}
= \frac{K_B}{2\Delta}.
\]

\textbf{Case $K_B > \Delta$:}

\[
P(\Omega_1) = \int_0^{\Delta/K_B} \frac{\gamma K_B}{\Delta}\, d\gamma
+ \int_{\Delta/K_B}^{1} 1\, d\gamma,
\]

\[
= \frac{1}{2}\frac{\Delta}{K_B} + \left(1 - \frac{\Delta}{K_B}\right)
= 1 - \frac{\Delta}{2K_B}.
\]

If, $\Delta= 800$, and $K_{B} \leq \Delta$ we have 

\[ P(\Omega_1) = \frac{1}{800} \int_{0}^{1} \int_{0}^{\gamma K_B} 1 \, d\delta \, d\gamma = \frac{1}{800} \int_{0}^{1} \gamma K_B \, d\gamma = \frac{K_B}{1600}. \]

where the capacities are found by solving Equation \ref{mathe} (using software for example,  Wolfram Mathematica).

Two stage multi-resource investment models are invaluable as they help 
mirror, energy systems. In the first stage capacity decisions are known, these are investment decisions and made before the demand or renewable availability is known. In the second second stage dispatch decisions, are made after uncertainty is realized.
The two‑stage structure also reveal how capacity choices $K_{A}$ and $K_{B}$ influence the probability of operating in different regions: fully renewable supply, diesel‑supported supply, or shortage. This leads to solutions that are more robust, economically efficient, and align with how real systems behave when facing variability and risk.

\section{Battery Energy Storage System and its effect in Electricity Generation}

Another method for maximizing the profit and minimizing the cost from electricity generation while reducing carbon emissions is
through battery energy
storage systems.
Moreover, an optimal sized battery energy storage system can help maximize the
benefits of the power both generated from PV systems and diesel sources while being financially economical. In this section, we propose a mixed-integer programming approach which decides  hour-by-hour to charge/discharge in order to maximize profit, consistent with Alberta’s market participation rules. The optimization problem is developed in a similar manner
to that described in \cite{Shafieea}
\subsection{Data}

\begin{figure}[H]
\includegraphics[width= 15cm]{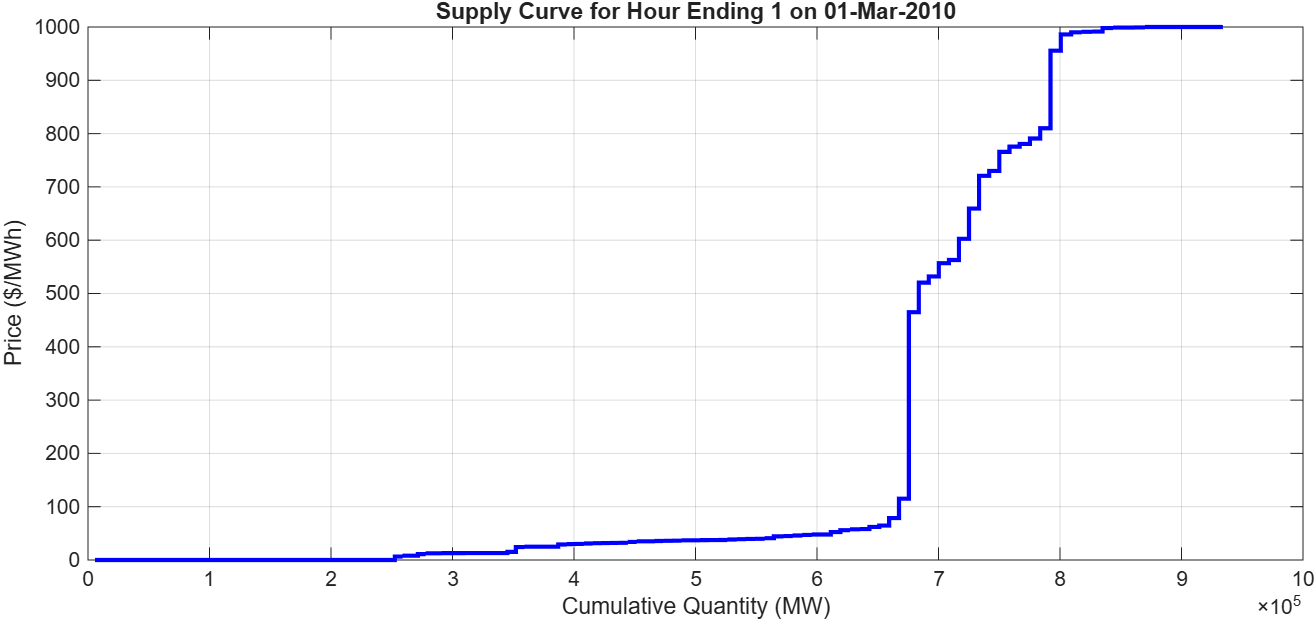}
  \caption{Bidstack on March 1st 2010}
  \label{fig:march}
\end{figure}

The data is in the form of a csv file indicating Day, Hour Ending, Price Quantity and Dispatched.
The market in Alberta has a price cap of CAD \$1000, and a price floor of \$0.
Producers are obliged to offer all of their available capacity into the market, but are under
no obligation to ensure that prices correspond to variable cost.
An example bid stack, from March 1 2010 is shown in Figure \ref{fig:march}.

As demonstrated in Figure \ref{fig:march} the shape of this
curve varies from hour to hour and the steps represent the price offered by generators for a quantity of power in a particular hour. The flat lines reflect cheaper prices however between 600,000-700,000MW and a vertical line indicating a drastic increase in the price.

\begin{figure}[H]
  \centering
  \begin{minipage}[b]{0.4\textwidth}
    \includegraphics[width=\textwidth]{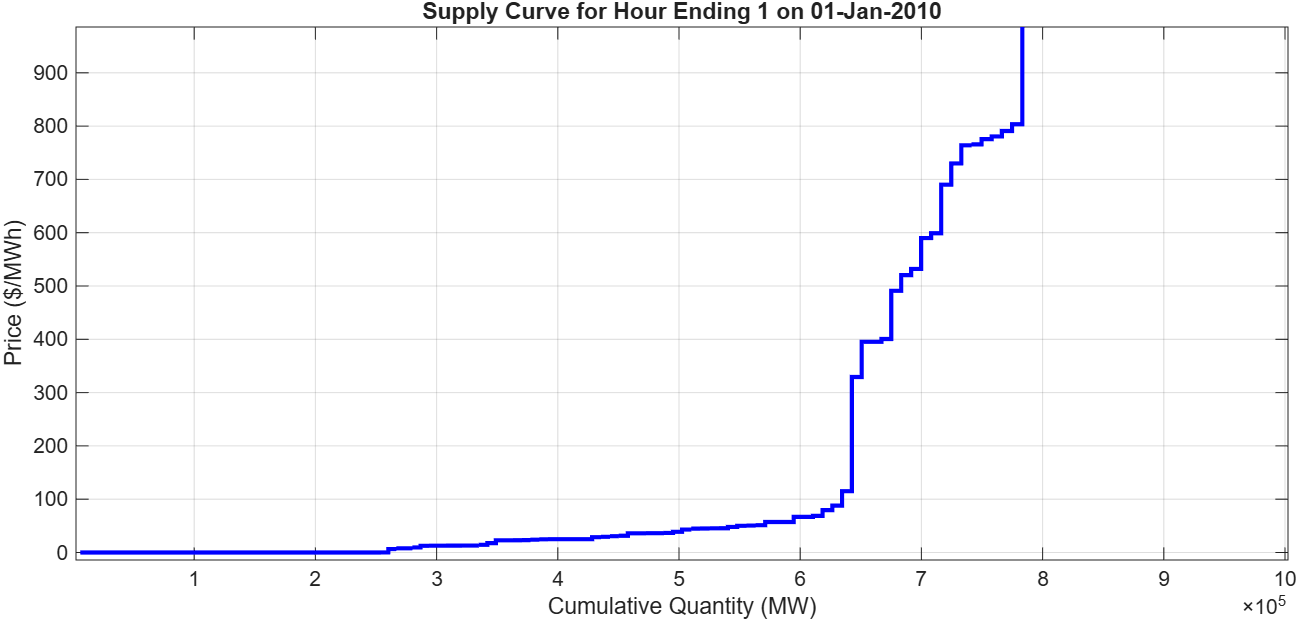}
    \caption{January 1  hour ending 1}\label{fig:jan1}
  \end{minipage}
  \hfill
  \begin{minipage}[b]{0.4\textwidth}
    \includegraphics[width=\textwidth]{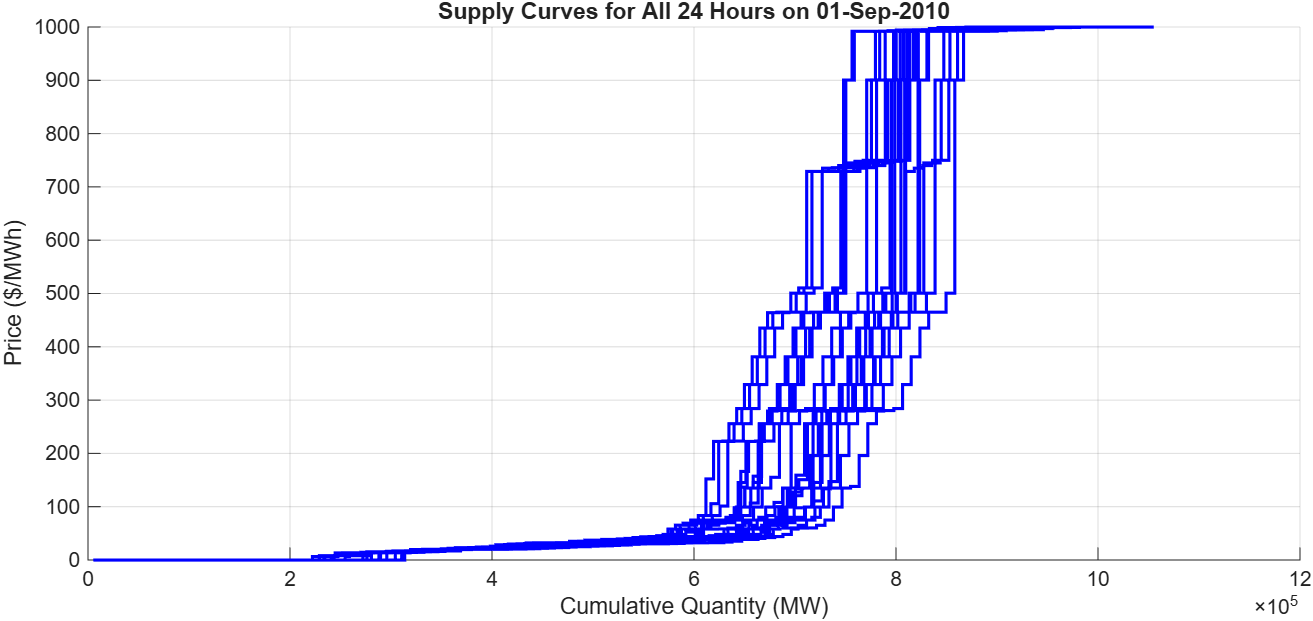}
    \caption{September 1, 24 hour}\label{fig:all}
  \end{minipage}
\end{figure}

Figures \ref{fig:jan1} and \ref{fig:all} show how the electricity prices vary throughout the day on January 1st, hour ending 1 and for all 24 hours throughout the day on September 1 2010 respectively. Not surprisingly, price increases with quantity as higher demand requires more expensive sources of electricity.
There's a notable steep price jump between roughly 600,000 MW and 800,000 MW, suggesting that during these hours, the grid needed to tap into much costlier electricity sources to meet demand.
Beyond 800,000 MW, prices often exceeded \$500/MWh and even approaching \$1000/MWh, indicating extreme scarcity or reliance on expensive peaking power plants. As shown in Figure \ref{fig:all} each blue curve represents one hour of the day and the shape and spread of the curves imply variability in supply conditions across the day, also indicating that the market participants submit their offers strategically depending on the market conditions.

\begin{figure}[H]
\includegraphics[width= 15cm]{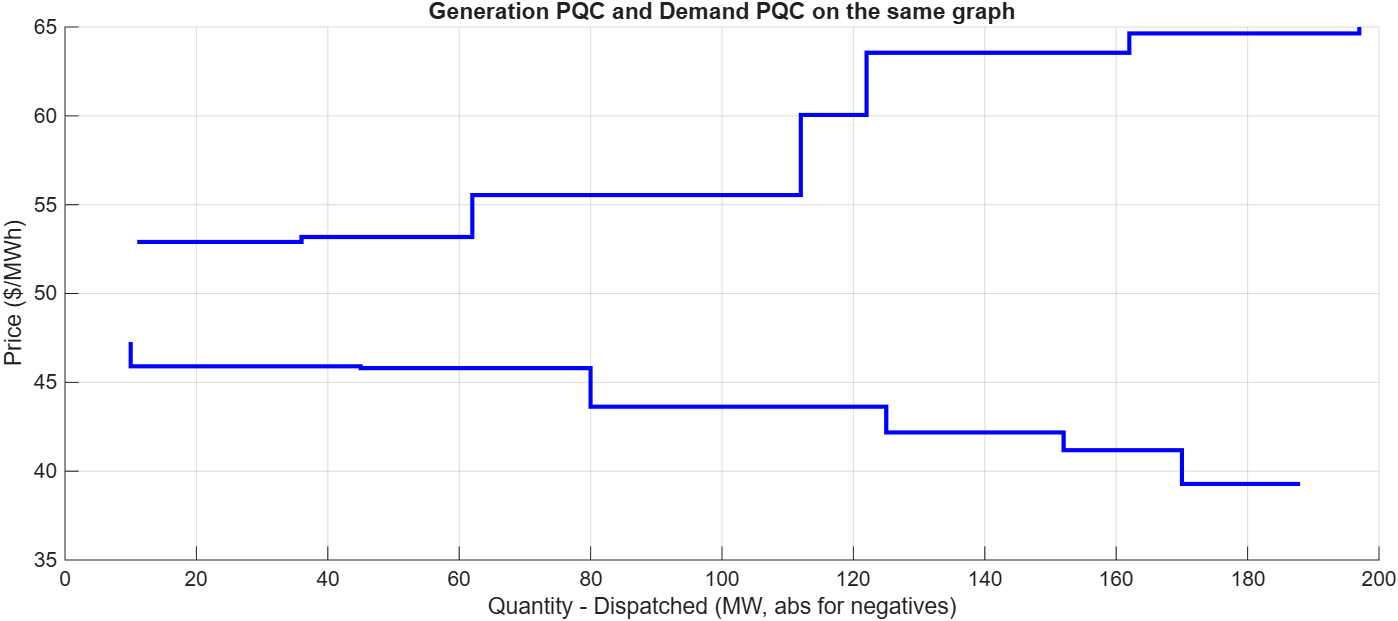}
  \caption{Hour ending 20- Charging and Discharging Blocks}
  \label{fig:DPQCs}
\end{figure}

Figure \ref{fig:DPQCs} shows the GPQC and DPQC for hour ending 20 on January 1st 2010. For the GPQC each horizontal segment represents a block of generation capacity offered at a fixed price. This is a supply curve and as more MW are dispatched, the price increases in steps, this reflects marginal cost bidding by generators.
The DPQC is a demand curve and shows how much consumers bid for electricity in blocks, with higher willingness to pay for initial MW and lower for additional consumption. 

\subsection{Battery Energy Storage Systems}

Using GPQCs and DPQCs as input data we develop  we develop a self-scheduling formulation in which the price is a variable and is a function of the charging or discharging quantities. Next, using this framework we  determine bidding strategies and assess the economic feasility of the energy storage systems providing opportunities for energy arbitrage in the Alberta electricity market.

\subsection{Optimization Model}

\textbf{List Of symbols}

\begin{tabular}{ll} $d$ & Day index, $d=1,\ldots,D$ \\ $h$ & Hour index, $h=1,\ldots,H$ \\ $s$ & Charge/discharge block index \\ $D$ & Set of days \\ $H$ & Set of hours \\ $S$ & Set of piecewise-linear blocks \\ \end{tabular}

\textbf{Decision Variables:}
\begin{align*}
x_{d,h,s}^d &\in \{0,1\} && \text{binary, select discharge block $s$ at $(d,h)$} \\
b_{d,h,s}^d &\geq 0      && \text{continuous, discharge quantity within block $s$ at $(d,h)$} \\
x_{d,h,s}^c &\in \{0,1\} && \text{binary, select charge block $s$ at $(d,h)$} \\
b_{d,h,s}^c &\geq 0      && \text{continuous, charge quantity within block $s$ at $(d,h)$} \\
\mathrm{SoC}_{d,h} &\in [0,400] && \text{state of charge at $(d,h)$}
\end{align*}

\vspace{0.3cm}

\textbf{Parameters:}
\begin{align*}
p_{d,h,s}    && \text{price of discharge block $s$ at $(d,h)$} \\
pc_{d,h,s}   && \text{price of charge block $s$ at $(d,h)$} \\
q_{d,h,s}    && \text{quantity of discharge block $s$ at $(d,h)$} \\
qc_{d,h,s}   && \text{quantity of charge block $s$ at $(d,h)$} \\
OC           && \text{operating cost per hour} \\
\eta         && \text{charge efficiency (e.g., 0.7)} \\
\mathrm{SoC}_0 && \text{initial state of charge (e.g., 100 MWh)}
\end{align*}

\subsection*{The Objective Function and Constraints }
Maximize net profit:
\begin{align}
\max \quad & 
\sum_{d \in D} \sum_{h \in H} 
\Bigg[
\sum_{s \in S_d(d,h)} p_{d,h,s} \big(b_{d,h,s}^d + x_{d,h,s}^d \cdot q_{d,h,s}^{\min}\big) \\
& \quad - \sum_{s \in S_c(d,h)} pc_{d,h,s} \big(b_{d,h,s}^c + x_{d,h,s}^c \cdot qc_{d,h,s}^{\min}\big) 
- OC
\Bigg]
\end{align}

\subsection*{Constraints}

\paragraph{}
\begin{align}
b_{d,h,s}^d &= f_{d,h,s}^d \cdot q_{d,h,s}^{\max}, \quad \forall d,h,s \in S_d(d,h) \\
b_{d,h,s}^c &= f_{d,h,s}^c \cdot qc_{d,h,s}^{\max}, \quad \forall d,h,s \in S_c(d,h)
\end{align}

\paragraph{}
\begin{align}
b_{d,h,s}^d &\leq x_{d,h,s}^d \cdot q_{d,h,s}^{\max}, \quad \forall d,h,s \in S_d(d,h) \\
b_{d,h,s}^c &\leq x_{d,h,s}^c \cdot qc_{d,h,s}^{\max}, \quad \forall d,h,s \in S_c(d,h)
\end{align}

\paragraph{}
\begin{align}
\sum_{s \in S_d(d,h)} x_{d,h,s}^d + \sum_{s \in S_c(d,h)} x_{d,h,s}^c \leq 1, 
\quad \forall d,h
\end{align}

\paragraph{}
\begin{align}
\text{SoC}_{d,h} &= \text{SoC}_{d,h-1} 
+ \eta \sum_{s \in S_c(d,h)} \big(b_{d,h,s}^c + x_{d,h,s}^c \cdot qc_{d,h,s}^{\min}\big) \\
& \quad - \sum_{s \in S_d(d,h)} \big(b_{d,h,s}^d + x_{d,h,s}^d \cdot q_{d,h,s}^{\min}\big),
\quad \forall d,h
\end{align}

with initial condition:

\[
\text{SoC}_{d=1,h=1} = \text{SoC}_0
\]

and bounds:

\[
0 \leq \text{SoC}_{d,h} \leq 400, \quad \forall d,h
\]

\begin{table}[htbp] 
\centering 
\caption{Battery Operation Schedule for 2010-01-01} \label{tab:battery_sch} \begin{tabular}{|c|c|c|c|c|}
\hline
Hour & SoC (MWh) & Mode & Electricity (MWh) & Price (\$/MWh)
\\ \hline
01 & 36.00 & Discharge & 64.00 & 57.25 \\ 
02 & 0.00 & Discharge & 36.00 & 57.24 \\ 03 & 0.00 & Idle & 0.00 & 0.00 \\
04 & 0.00 & Idle & 0.00 & 0.00 \\ 05 & 100.00 & Charge & 142.86 & 28.73 \\ 06 & 0.00 & Discharge & 100.00 & 45.00 \\ 07 & 0.00 & Idle & 0.00 & 0.00 \\ 08 & 154.70 & Charge & 221.00 & 16.40 \\ 09 & 85.70 & Discharge & 69.00 & 29.56 \\ 10 & 286.60 & Charge & 287.00 & 14.60 \\ 11 & 400.00 & Charge & 162.00 & 15.15 \\ 12 & 400.00 & Idle & 0.00 & 0.00 \\ 13 & 304.00 & Discharge & 96.00 & 43.97 \\ 14 & 304.00 & Idle & 0.00 & 0.00 \\ 15 & 400.00 & Charge & 137.14 & 29.56 \\ 16 & 400.00 & Idle & 0.00 & 0.00 \\ 17 & 400.00 & Idle & 0.00 & 0.00 \\ 18 & 312.00 & Discharge & 88.00 & 66.86 \\ 19 & 216.00 & Discharge & 96.00 & 55.06 \\ 20 & 146.00 & Discharge & 70.00 & 57.19 \\ 21 & 31.00 & Discharge & 115.00 & 49.00 \\ 22 & 31.00 & Idle & 0.00 & 0.00 \\ 23 & 31.00 & Idle & 0.00 & 0.00 \\ 24 & 0.00 & Discharge & 31.00 & 47.91 \\ \hline 
\end{tabular}
\end{table}

\begin{table}[htbp] \centering \caption{Charge and Discharge Block Details for 1 January 2010} \label{tab:block_details} 
\begin{tabular}{|c|c|c|c|c|} \hline Hour & Mode & Block & Electricity (MWh) & Price (\$/MWh) \\ \hline 01 & Discharge & 2 & 64.00 & 57.25 \\ 02 & Discharge & 2 & 36.00 & 57.24 \\ 05 & Charge & 1 & 142.86 & 28.73 \\ 06 & Discharge & 2 & 100.00 & 45.00 \\ 08 & Charge & 2 & 221.00 & 16.40 \\ 09 & Discharge & 2 & 69.00 & 29.56 \\ 10 & Charge & 2 & 287.00 & 14.60 \\ 11 & Charge & 1 & 162.00 & 15.15 \\ 13 & Discharge & 2 & 96.00 & 43.97 \\ 15 & Charge & 1 & 137.14 & 29.56 \\ 18 & Discharge & 2 & 88.00 & 66.86 \\ 19 & Discharge & 2 & 96.00 & 55.06 \\ 20 & Discharge & 2 & 70.00 & 57.19 \\ 21 & Discharge & 2 & 115.00 & 49.00 \\ 24 & Discharge & 2 & 31.00 & 47.91 \\ \hline 
\end{tabular} 
\end{table}

According to Table (\ref{tab:battery_sch}), the battery supplies energy during its discharging hours, for example during hours ending 1, 2 and 6. During these hours, the prices decrease, benefitting customers. However during the battery's charging hours the battery the prices increase. Table (\ref{tab:block_details}) shows the block details for the optimization problem with the profit or Gurobi objective value was found to be $\$8351.22$

\newpage

\section{Dynamic Programmming towards the Battery Storage Problem}

The novelty in this research is two-fold. First we provide a comparison of statistical and optimization methods for maximizing profit while minimizing cost and reducing carbon emission in electricity generation.  Next we propose a novel approach based on stochastic dynamic programming towards the battery storage problem for electricity generation. To our  knowledge, such a comparison has never been done before in this context.

Stochastic dynamic programming has become a powerful tool when applied to a variety of 
problems (see \cite{ Lawrence2, Wu2, Rahaman,  Lawrence3, Ross, Verweij} and the one-armed bandit processes was formulated as a stochastic process using a Bayesian approach and solved using dynamic programming by Wang and Bickis in \cite{WangBickis}.
A dynamic programming formulation towards battery energy storage operators have been studied in various research \cite{Jiang} . However these research is still different to the approach proposed in this paper. In \cite{Jiang} the authors propose a stochastic dynamic programming model that accounts for the amount of energy stored in a battery at the beginning and the end of an hour. The objective function maximizes the revenue function based on the resource/energy level, prices/bids and number of charge/discharge cycles and using approximate dynamic programming to solve the problem. programming towards the battery energy 

In this research, we adopt a SDP formulation towards the Battery Storage problem in order to optimize profit in an energy market. Our implementaiton  is as follows:

Let $J^{*}_{t}$ denote the maximum expected return
for a t-stage problem and assign vectors $\vec{x}_{t}$  as the state vector of the system on  state space  $\mathcal{X} $ and vector $\vec{a}_{t}$ as the control decision vector on action space $\mathcal{A}$.
Where s= $1,...n^{c}$- the number of charging blocks  and $s'={1,...n^{d}}$ the number of discharging blocks.

If we let $T$ denote the number of stages and $t$ denote the number of stages to go, the problem involves maximizing the following expected cost over a finite number of stages $T$.

\begin{equation}\label{optimal}
J_{K}:= E\left\{ c_{T}(x_{t}) + \sum_{t=1}^{T-1} c_{t} (\vec{x}_{t}, \vec{a}_{t})\right\}
\end{equation}

We observe that the optimal allocation to the problem $\max J_{t}$ given by,

\begin{equation}\label{opti}
\max_{\vec{u}_{1},..., {\vec{u}_{t}}} E\left\{ c_{T}(x_{T}) + \sum_{t=1}^{T-1} c_{t} (\vec{x}_{t}, \vec{a}_{t})\right\}
\end{equation}

To find the optimal allocation for the optimization problem $(\ref{opti})$, we start by defining value functions, $V_T$. The value functions $\left\{V_t\right\} $  are a sequence of functions $V_{t}
: \mathcal{X} \rightarrow R$, defined by the
following: 

\begin{enumerate}

\item At the last stage $T$, $V_{T} (x) = c_T (x), \forall x$.

\item At time $k = T-1, . . . ,0$, $Q_t: \mathcal{X} \times \mathcal{U} \rightarrow R$, the $Q$ function is defined by

\begin{equation*}
Q_t(x, u) = c_t(x, u) + E(V_{t+1}(X_{t+1}))
\end{equation*}

where $Q_t(x, u)$ is the expected cost or reward in regard to states $x$ and action $u$.

\item At time $t$, the value function is 

\begin{equation*}
V_{t}(x) = \max_{u \in U} Q_t(x, u)
\end{equation*}

Therefore the optimal policy is given by 

\begin{equation}\label{optimality}
g^*_{t}= \operatorname*{arg\,max}_{u\in \mathcal{U}} Q_t(x,u)
\end{equation} 

\end{enumerate}

The optimality of Equation \ref{optimality} is illustrated  if we define the cost-to-go function by

\begin{equation*}\label{optimal}
J_{t}(x,g):= E\left\{c_{T}(x_{T}) + \sum_{s=t}^{T-1} c_{s} (\vec{x}_{s}, \vec{u}_{s})\right\}
\end{equation*} 

then $J_{t}(x;g) \leq V_{k} (x)$

where the equality holds if

\begin{equation*}\label{optimal}
g_{v}= \operatorname*{arg\,max}_{u\in \mathcal{A}} Q_v(x,a)
\end{equation*} 

$\forall s \geq k$.

The system dynamics depends on the state-action
pair
$(x_{t}, u_{t})$ at current stage $t$ and probabilities $p_t$:

\begin{equation}
x_{t-1} = f_t(x_{t}, u_{t}, p_{t}), \text{for  t = 0, 1, . . .$T-1$}
\end{equation}
where $P(x_{t-1}| x_{t}, u_{t},p_{t})$.

The Bellman recursion is then

\begin{equation*}
Q_t(x, u) = c_t(x, u) + E(V_{t+1}(X_{t+1}))
\end{equation*}

\subsection{An Implementation of Stochastic Dynamic Programming to Battery Storage problems}

In this section, we propose a stochastic dynamic programming approach to model a battery energy storage system and maximize profit in electricity generation. Energy markets, particularly electricity markets inherently have probabilistic prices due to volatile renewable energy, demand shift and grid congesting, requiring sophisticated probabilitic forecasting, hedging and optimal trading strategies. Methods for predicting the uncertainty in prices vary and include Polynomial mapping and Deep Learninng methods. This research proposes a novel SDP approach to account for the uncertainty in prices associated with the while maximizing profit and revenue in electricity generation.

\subsection*{Stochastic Dynamic Programming Model }

For finite horizon stochastic problems, the optimality equations at time $t$ are given by:

\[
V(x_t) = \max_{u_t \in U} 
\mathbb{E}\!\left( C_T(x_T, u_t) + \sum_{t=1}^{T-1} V_{t+1}(x_t, u_t, w_t) \right)
\]

which is equivalent to solving the following, known as Bellman’s optimality equations:

\[
V(x_t) = \max_{u_t \in U} 
\Big( C_t(x_t, u_t) + \mathbb{E}\big[ V_{t+1}(x_t, u_t, w_t) \big] \Big)
\tag{1.5}
\]

where $x_t$ is the state variable, in this case the state of charge and the price at time $t$, and summarizes past information useful for solving future states, $u_t$ is the control or decision variable, that is whether to charge or discharge , and $w_t$ is a random parameter capturing the uncertainty  of the prices

\vspace{3mm}

Given the value function \(V_t(e)\) as the optimal expected value, then the optimization problem with \(t\) stages remaining can be written as:

\[
V_t(e) \;=\; \max_{u \in \mathcal{U}_t(e)} \left\{ \mathbb{E}\big[\Pi_t(u)\big] \;+\; V_{t-1}\!\big(e') \right\}, 
\qquad t=1,\dots,T,
\]
where, 
\[
\big(e'= e + \eta_c (b)^{ch} - b^{dis}\big) 
\]

and 

\[
\Pi_t(u_t) \;=\; 
 p_{t,s} \times b^{dis}_{t}  - pc_{t,s} \times   b^{ch}_{t}  
\]

with the last stage as:

\[
V_0(e) \;=\; c_T(e) \quad \text{i.e., } c_T(e) = 0 
\]

\subsection*{Example without incoporating the stochastic nature of prices}
Consider a toy example involving t=2 hours as the time given and s= 2 blocks. 
The discharging and charging price per hour  at time $t$ and  step $s$ is given by $p^{dis}_{t,s,}$ and $p^{ch}_{t,s}$    respectively. The corresponding values are   $p_{1,1}^{dis} = \$20$, $pc_{1,2}= \$10 $,  $p_{2,1} = \$25$, $pc_{2,2}= \$5 $ with initial conditions, $V_{0} (E) =0$, $E(0)=5$. The reward function $r_{t}$ is calculated as $r_{t} = p_{t,s} \times b^{dis}_{t}  - p^{ch}_{t,s} \times   b^{ch}_{t} $.

For time $t=1$, suppose the battery is in state $E_0=5$ and $u_1=\text{Idle}$ that is, neither charging nor discharging. Using $b_{1,1}^{ch}=0$ and $b_{1,1}^{dis}=0$.Therefore,

\[\centering
    \begin{aligned} 
      V_{1} (5)    &= r_{1} + V_{0} (5) = 0
    \end{aligned}
    \]

\vspace{5mm}
For time t=1, and $u_{t}$= Discharging 1 MWh $ b^{dis}_{1,1} $ = 1, $b^{ch}_{1,1} $ = 0, $ E_{1} = 5-1 =4,  r_{1} = 20 *1 =20$

Therefore, 
\[\centering
    \begin{aligned}
      V_{1} (4)    &= r_{1} + V_{0} (5) = 20
    \end{aligned}
    \]

\vspace{5mm}

 For time t=1, and $u_{1}$ = charging, $E^{dis}_{0}$ = 5,  $b^{ch}_{1,1} $ = 0,$ b_{1,1} $ = 1, $E_{1} = 5 + 0.8 *1 = 5.8, r_{1} $   =-10 * 1  =-10,    $E_{1} = 5 + 0.8 *1 = 5.8$

\[\centering
    \begin{aligned}
      V_{1} (5.8) = r_{1} + V_{0} (5.8) = -10
    \end{aligned}
    \]

The maximum value is found by evaluating all possible feasible actions.  That is, all possible discharge and charge amounts over all possible values and outcomes. The  dynamic programming model can also incorporate the stochastic nature of prices as done in several papers (see for example \cite{Now} and \cite{Jiang}).

\section{Discussion}
The results of Section 3 indicate the probit and Bayesian model suited for determining the probability of a mini-grid project being successful and handling probabilities.
 The probit model
captures the significant relationship between project success and its covariates,
Community Ownership and Storage and its predictive performance is closely related
to the Bayesian model. Bayesian methods are useful when the data is scarce and
particularly for handling latent variables, such as the probability of a mini-project
being successful in this case.

Sections 4 focuses on the mathematical optimization model for minimizing cost in generating electricity while reducing carbon emissions. In particular,
a two‑stage multi-resource investment model is described in Section 4 where capacities are chosen before the uncertainty is revealed. In the first stage, investment decisions are made for different types of capacities—typically a reliable but costly resource, for example diesel and a cheaper renewable resource, whose availability is uncertain. In the second stage, after the demand and availability are realized, resources are allocated to meet the demand, and a penalty is paid for any unmet demand. The model naturally divides the demand space into different regions. The expected cost combines the probability of each region when the renewables alone meet the demand, when the renewables alone are insufficient and diesel is used and when diesel and renewables are still insufficient to meet the demand. This captures  the trade‑off between upfront  capacity investments for diesel and renewable purchases $(K_{A}$ and  $K_{B})$ and the chance that, later on, demand is high or availability is low, forcing expensive penalty costs.

Sections 5 and 6 introduce battery energy storage systems as another method for  maximizing profit and minimizing the cost in electricity
generation while reducing carbon emissions. In Section 5, we propose a mixed-integer programming
approach which  decides hour-by-hour to charge or discharge in order to maximize
profit, consistent with Alberta’s market participation rules.  The battery successfully executes a daily arbitrage
strategy, that is, charging when prices are relatively low and discharging when prices are
high. The battery’s value is not
just in daily smoothing, but in capturing rare price spikes and dampens this spike by discharging during peak price hours. Thus, consumers
pay less, and the system benefits from price stability.

Section 6 implements a dynamic programming framework for optimizing battery storage operations in which the uncertainty  in electricity prices can be incorporated. The state variable $x_t$ (or state of charge $e_t$) summarizes all relevant past information, the control variable $u_t$ represents charging or discharging decisions, and $w_t$ captures price uncertainty.
The volatility of electricity prices for the SDP model can be evaluated using different methods. In this paper, we illustrate the effect of a dynamic programming model applied to a battery energy storage system and leave the stochastic component of prices to be incorporated in future research 

\section{Conclusion}

This paper introduces a statistical and optimization modeling framework to examine decision‑making strategies in modern energy systems. 
First, a Bayesian hierarchical model is compared with a probit model on 104 mini‑grids installed across the globe to highlight the strengths of statistical modeling and provide insights into the success of mini-grid projects in electricity generation. Second, a deterministic mixed-integer optimization model is developed which maximizes profit while reducing carbon emissions. Results indicate that battery systems can execute effective arbitrage strategies that is,   charging when prices are relatively low, discharging when prices are high. 

Furthermore, we propose a dynamic programming modeling framework to the battery energy storage system. Preliminary results indicate that dynamic programming method offers a promising alternative to mixed‑integer programming, particularly in volatile electricity markets. In a follow-up paper we will explore methods for capturing the volatility and provide extensive experimental results.

The results highlighted in this paper emphasize the importance of integrating statistical and optimization methods towards energy systems. By accounting for uncertainty, these frameworks support more resilient decision-making efforts.
Future research may entail  incorporating further statistical, probabilistic and optimization based methods towards extensive research in developing effective, clean, reliable energy systems.

\textbf{Acknowledgments}

I gratefully acknowledge that this research was supported in  part by the Pacific Institute for the Mathematical Sciences. Many thanks also to my postdoctoral Supervisors at the University of Calgary for the data and their guidance during this time.

\begin{appendices}




\end{appendices}


\newpage

\bibliography{sn-bibliography}

\end{document}